\documentclass{article}

\usepackage{PRIMEarxiv}

\usepackage[utf8]{inputenc} 
\usepackage[T1]{fontenc}    
\usepackage{hyperref}       
\usepackage{url}            
\usepackage{booktabs}       
\usepackage{amsfonts}       
\usepackage{nicefrac}       
\usepackage{microtype}      
\usepackage{lipsum}
\usepackage{fancyhdr}       
\usepackage{graphicx}       
\graphicspath{{media/}}     
\usepackage{amsmath} 
\usepackage{float} 
\usepackage[numbers,sort&compress]{natbib}

\title{NETWORK-LEVEL TRAVEL TIME PREDICTION CONSIDERING THE EFFECTS OF WEATHER AND SEASONALITY}

\author{
  Yufei Ai \\
  Department of Civil \& Environmental Engineering \\
  University of Houston \\
  \texttt{yufeiai91@gmail.com} \\
   \And
  Yao Yu, Ph.D. \\
  Meta Inc. \\
  \texttt{yuyao08@gmail.com} \\  
   \And
  Wenjing Pu, Ph.D. \\
  Boston Consulting Group\\
  \texttt{Pu.Wenjing@bcg.com} \\  
   \And
  Lu Gao, Ph.D. \\
  Department of Civil \& Environmental Engineering \\
  University of Houston \\
  \texttt{lgao5@central.uh.edu} \\
   \And
  Yihao Ren \\
  Department of Transportation, Logistics and Finance \\
  North Dakota State University \\
  \texttt{yihao.ren@ndsu.edu} \\  
}

\begin{document}
\maketitle

\begin{abstract}
Accurately predicting travel time information can be helpful for travelers. This study proposes a framework for predicting network-level travel time index (TTI) using machine learning models. A case study was performed on more than 50,000 TTI data collected from the Washington DC area over 6 years. The proposed approach is also able to identify the effects of weather and seasonality. The performances of the machine learning models were assessed and compared with each other. It was shown that the ridge regression model outperformed the other models in both short-term and long-term predictions.
\end{abstract}


\section{Introduction}

Researches on predicting travel time can be divided into three categories by the time scale of prediction: long-term, short-term, and real-time travel time prediction. The generally adopted methods can be categorized into parametric (linear regression models and time series analysis) and nonparametric approaches (neural networks and pattern searching) \cite{Oh2015ShortTermReview}.

\subsection{Real-Time Prediction}

Real-time travel prediction is aimed at providing the most up-to-date traffic information to the drivers with the input of GPS information which can be retrieved periodically to the server. \citet{Wisitpongphan2012MLFFANN} conducted a study based on the GPS data of 297 probing vehicles driving on a 22km-long road section. The study used artificial neural network (ANN) model with multi-layer feed forward algorithm. Given the time and vehicles’ coordinate, speed, and heading, their model is able to predict the travel time on any given road in real-time. The predictions during non-rush hour periods are very close to the actual results whereas during rush hours, it still remains unstable due to the traffic congestion. Moreover, significant effects of different weekday and time of the day were observed in the data sets. \citet{GurmuFan2014BusTTIGPS} developed a multilayer perceptron (MLP) artificial neural network (ANN) model to predict bus travel time at a given downstream bus stop. They used arrival and departure times at each bus stop of the selected buses as input and obtain their best accuracy with the error of less than 10\%. The authors pointed out that ANN models show advantages on its good robustness and capacity of predicting complicated models while it contains some unknown, highly nonlinear relationship, or noisy information in the data sets. However, the data size has to be large enough to ensure the accuracy which could be an obstacle for data acquisition in practical.

\subsection{Short-Term Prediction}

A considerable number of researches has been done by many researchers on the subject of short-term travel time forecasting. In these researches travel time at a given period of time in the future is predicted based on the historical and current traffic data. The most common used methodologies include: (1) linear regression models
\cite{KwonCoifmanBickel2000DayToDayTrends,RiceVanZwet2004SimpleEffective}; (2) auto-regressive integrated moving average (ARIMA) models \cite{SaitoWatanabe1995PredictionDissemination,Oda1990VehicleSensorPrediction}; (3) artificial neuron networks (ANN) models \cite{HuiskenVanBerkum2002ShortTermNN,Liu2021HighwayNetworkDL}. \citet{RiceVanZwet2004SimpleEffective} utilized a simple linear regression model with time-varying coefficients. The model was proposed in \cite{HastieTibshirani1993VaryingCoefficient} to perform travel time predictions on a given section of free way. Travel time was predicted for a certain time lag ahead departure; given the historical average travel time data in all the periods in previous 34 weekdays, and the instantaneous travel time data obtained from multiple types of measurement devices, such as single-loop detector or probe vehicles, the model was tested to be valid for the time lag ranging from 0 to 60 minutes. \citet{WuHoLee2004SVRTravelTime} applied support vector progression model for predicting travel time for high ways near Taipai and analyzing the daily and weekly patterns. The accuracy measures in terms of relative mean errors (RME) and root-mean-squared errors (RMSE) were shown to be within 5 to 8\%. The satisfying results were believed to be credited with SVR’s great generalization ability and guaranteed global minima of the results whereas only local minima can be located with neural networks. \citet{Oda1990VehicleSensorPrediction} predicted the travel time with ARIMA model, in which the travel time observations were arranged in stationary time series and decomposed into three termsan auto-regressive term (AR), a moving-average term (MA), and an integrating term that accounts for the data noises. The prediction was completed through minimizing the noise term. The best accuracy was achieved for the morning peak data where the error was controlled within 6\%. \citet{SaitoWatanabe1995PredictionDissemination} predicted travel time 60 minutes ahead with last 30 minutes’ observations. The average error of the prediction results is less than 3 minutes apart from actual values. As a non-parametric approach, neural networks also play an important role in the research on short term travel time prediction. \citet{HuiskenVanBerkum2003ComparativeShortRange} proposed a multilayer feed forward based ANN model for predicting travel time at a given highway and compared its performance with the results from two naive methods: dynamic travel time estimation (DTTE) and static travel time estimation (STTE). The model was trained by the flow and speed data on a one-minute basis. The information was collected from inductive loop detectors at 21 locations on a given freeway section, at a corresponding time of the day. It was found that ANN models outperform the two naive models significantly. In order to extend the travel time prediction model to road traffic networks rather than a single road section, \citet{WangTsapakisZhong2016STDelayNN} integrated space-time neuron into the neural networks by adding a hidden layer. The road traffic network was represented numerically in the form of a spatial weight matrix. Performance of the space-time delay neural networks was compared against and proved to be better than the other three models, Naive, ARIMA, and space-time ARIMA, at the 5, 15 and 30 minutes forecasting horizons. \citet{Moreira2009VaryingParametersBus} did a series of works to predict travel time in a three-day prediction horizon based on a 244-day data set. They utilized and compared three non-parametric regression methods in R package: Projection Pursuit Regression (PPR), Support Vector Machine (SVM) and Random Forest (RF). Departure time, weekday, day of the year, and day type (holiday, bridge day, etc.) were found to be the most valuable independent variables. In their following work \cite{MendesMoreira2015HeterogeneousEnsembles}, study on 128 combinations which composed of the three models and respective parameter sets was done for each of the six studied routes by using a heterogeneous ensemble approach with dynamic selection. The results showed that the ensemble approach is able to increase accuracy by 8.2\% against the use of a single algorithm and parameter set.

\subsection{Long-Term Prediction}

\citet{Klunder2007LongTermHistorical} conducted a long-term travel time prediction by using K-nearest neighbor (K-NN) approach. This is a typical pattern searching pattern approach. As a non-parametric method, it makes predictions by searching out the traffic patterns on similarities among data instead of defining parameters and/or distributional assumptions on input and output variables. Case study was done on a single route of three 25-km successive motorways with date and travel time data for every 15 minutes. School holidays, major events, special days and accurate precipitation data were also considered. The prediction in non-rush hour (8:00am and 10:00pm), with the RME of the median 4.6\% and 3.3\%, perform better than that in rush hour (12:00pm and 5:30pm) 7.6\% and 19.4\%.

\subsection{Existing Research on Seasonality and Weather}

Numerous studies have investigated the role of seasonality in transportation systems \cite{gao2007using,MemmottYoung2008SeasonalCongestion,MoriKockelman2024SeasonalDemandVariations,KashfiBunkerYigitcanlar2015ComplexSeasonalityTransit, gao2008robust,KashfiBunkerYigitcanlar2016ComplexSeasonalityWeatherTransit,SinghalKamgaYazici2014WeatherUrbanTransitRidership,StoverMcCormack2012WeatherBusRidership,DoiAllen1986MonthlyRidershipSeasonalVariation,gao2011performance,Changnon1996SummerPrecipitationUrbanTransportation,Nankervis1999BicycleCommutingSeasonalVariation,gao2010effect,Splawinska2017SeasonalTrafficVolumes,GastaldiGeccheleRossi2014AADTOneWeekCounts,GastaldiRossiGeccheleDellaLucia2013AADTSeasonalCounts,Cai2011NonparametricSeasonalFactorsAADT,gao2012bayesian,ElkhoulyAlhadidiRakha2025ComplexSeasonalTrafficCounts,BartuskaHanzlLizbetin2021UrbanTrafficDetectorsSeasonality}. For example, \citet{Nookala2006WeatherImpact} used linear regression model with time-varying coefficients to conduct a short-time travel time prediction with an emphasis on investigating the weather impacts. The linearity between a set of weather indexes, traffic volume, and dynamics was investigated by measurement of correlation coefficients. Conclusion was drawn that the daily total traffic volume decreased under non-ideal weather indexes while the congestion increased as the freeway capacity dropped. The results of travel time prediction showed that the prediction errors become more significant under changing weather indexes. \citet{ElFaouzi2010MotorwayTTI} integrated weather effect investigation into their travel time prediction by using database search for similar profiles extraction based on toll collection data. Weather data analyzed in this work include rain, temperature, wind direction and speed. The data sets in this study consist of the whole hourly measurements in 43 days. It was shown that it provides more accurate predictions on travel time in junction with weather indexes. \citet{Klunder2007LongTermHistorical} indicated that the accuracies of predicted travel time were improved by including the accurate precipitation data into input variable set. RME of median decreased from 15\% to 7.6\% for the morning rush hour period and from 34.4\% to 19.4\% for the evening peak hours.

\section{Case Study}

In this research, hourly TTI data collected between Jan 1, 2010 and Jun 26, 2016 in the DC area and corresponding daily weather index data were used. 

\subsection{Descriptive Statistics}

In order to investigate data variation caused by effect of different time, effect of different weekday, effect of different season, and effect of different year, the mean historical TTI has been calculated under different time scale for descriptive statistical analysis. 

\begin{figure}[h]
    \centering
    \includegraphics[width=0.75\linewidth]{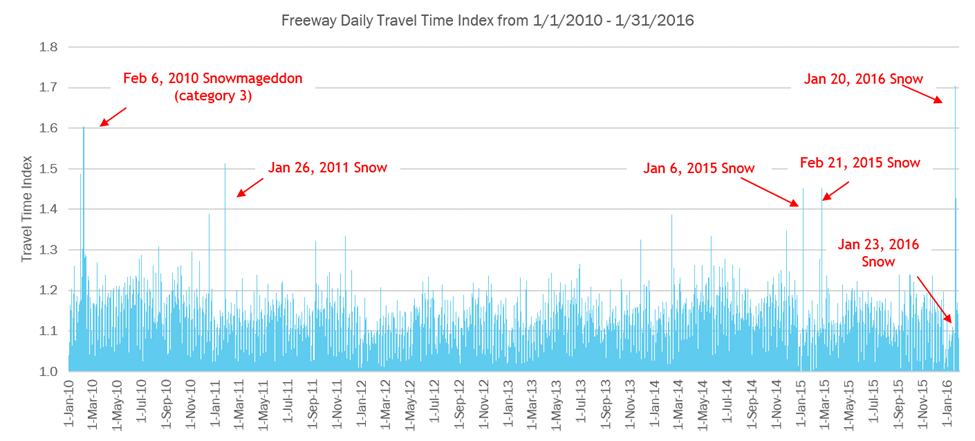}
    \caption{Daily Travel Time Index from 1/1/2010 to 1/31/2016}
    \label{fig:fig1}
\end{figure}

Figure \ref{fig:fig1} shows the daily–averaged TTI during the given time period. Further investigation has been done with regards to several significant peak values of TTI. As labeled in the figure, there exists a significant relationship between snow event and high value of daily averaged TTI. It can be seen that weather indexes, especially snow, play an important role to the significant high TTI values on a daily basis.

\begin{figure}[h]
    \centering
    \includegraphics[width=0.75\linewidth]{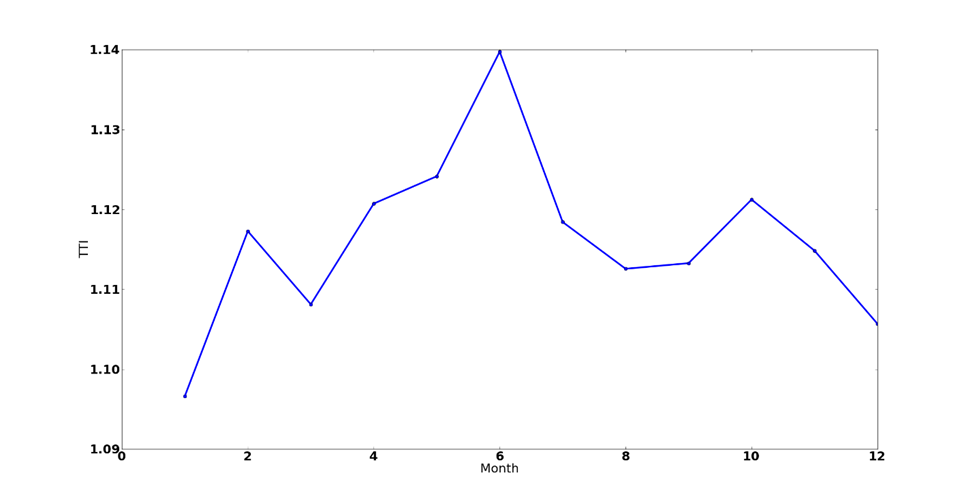}
    \caption{Averaged Travel Time Index as a Function of Month}
    \label{fig:fig2}
\end{figure}

However, the relationship between daily TTI and snow event does not lead to the conclusion that TTI values in winter is generally higher than in the other seasons. As shown in Figure \ref{fig:fig2}, the maximum monthly-averaged historical TTI occurs in June. May and October are also associated with relatively high TTI values. Significance of monthly effect is not observed from this study as the variation TTI over different month is small.

\begin{figure}[h]
    \centering
    \includegraphics[width=0.75\linewidth]{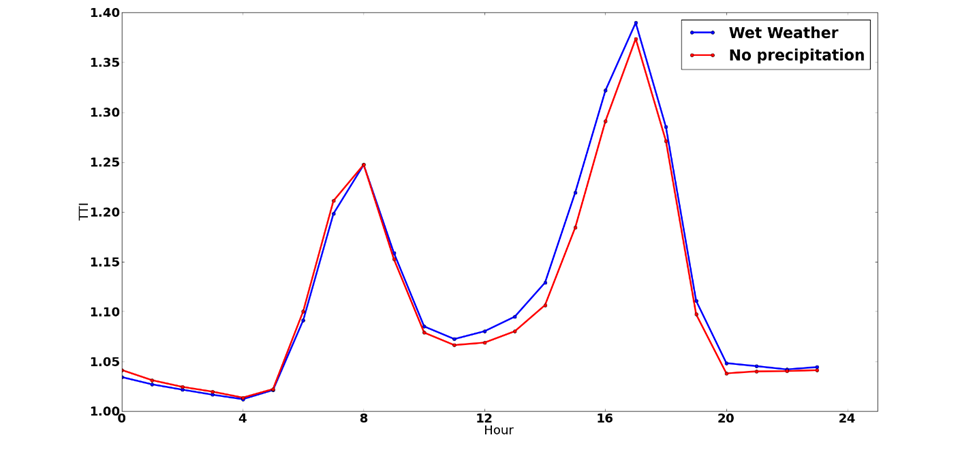}
    \caption{Averaged Travel Time Index over Time of a Day}
    \label{fig:fig3}
\end{figure}

Figure \ref{fig:fig3} shows the historical averaged TTI as a function of time. The red line is resulted from the data in the days with no precipitation, and the blue line is resulted in the days that has precipitation. It can be seen that TTI values reach the peak values at 08:00 and 17:00, which are within the morning and evening rush hours. The variation of average TTI in different time of a day is very significant. Therefore, it can be concluded that the hourly pattern is an important factor to be considered in data preparation. It can also be observed that the historical averaged TTI based on the data from wet days have slightly greater values than that from the days with no precipitation.

\begin{figure}[h]
    \centering
    \includegraphics[width=0.75\linewidth]{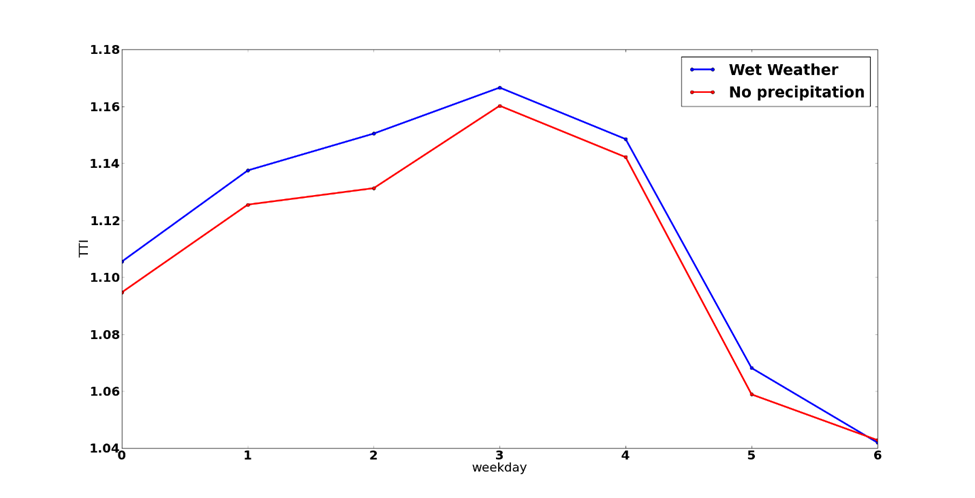}
    \caption{Averaged Travel Time Index per Weekday}
    \label{fig:fig4}
\end{figure}

Daily pattern is another significant factor in the study of historical data of TTI. The mean historical TTI is plotted against each weekday in Figure \ref{fig:fig4}, where 0 through 6 on X–axis represent Sunday to Saturday. As observed from the figure, TTI is generally highest on Wednesday and lowest on Saturday. TTI on weekdays are higher than weekends. It can also be observed that the historical averaged TTI on days with high precipitation are higher than TTI on days with low precipitation. This may be attributed to the traffic congestion caused by the wet weather.

\begin{figure}[h]
    \centering
    \includegraphics[width=0.75\linewidth]{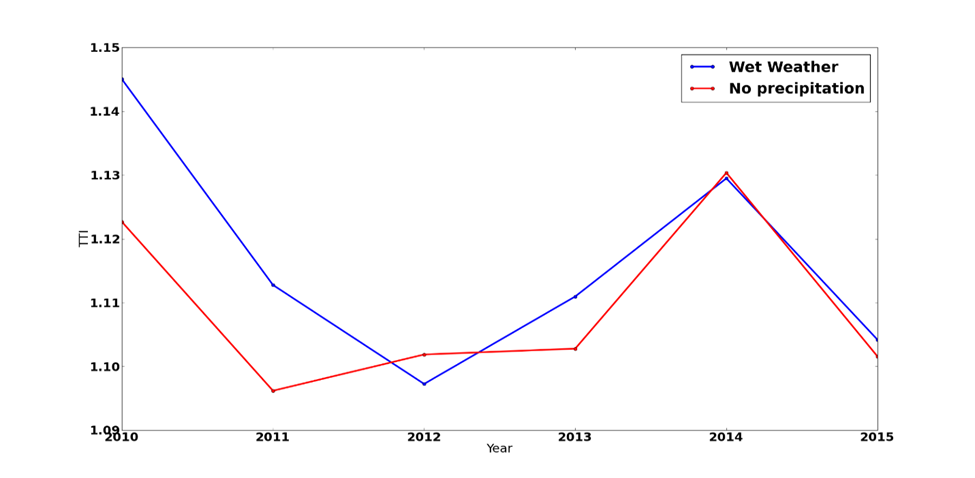}
    \caption{Averaged Travel Time Index per Year}
    \label{fig:fig5}
\end{figure}

The effect of different year is also studied by calculating the yearly-average TTI from 2010 to 2015. It is plotted in Figure \ref{fig:fig5}. No regular pattern has been witnessed from the figure. The relatively small variations indicate that the TTI levels are relatively stable during the years of interest.

\subsection{Model evaluation criterion}
In this study, the prediction of TTI is considered as a machine learning regression problem. For continuous-valued objects, there are varieties of regression algorithms designed to minimize the difference between the output from the predictor and the actual values. Therefore, the predictor can be used for future predictions with a promising accuracy. Coefficient of determination ($R^{2}$), named as score, is used as a measurement of the model accuracy. It is defined by Equations~(1)--(3).

\begin{align}
R^{2} &= 1 - \frac{SS_{\mathrm{res}}}{SS_{\mathrm{tot}}} \tag{1}\\
SS_{\mathrm{res}} &= \sum_{i}\left(y_i - f_i\right)^{2} \tag{2}\\
SS_{\mathrm{tot}} &= \sum_{i}\left(y_i - \bar{y}\right)^{2} \tag{3}
\end{align}

\noindent where $y_i$ is the actual value at testing/validation data point $i$; 
$f_i$ is the predicted value at testing/validation data point $i$; 
$\bar{y}$ is the mean of actual values of all testing/validation data points; 
$SS_{\mathrm{res}}$ is the residual sum of squares, as defined by Equation~(2); 
and $SS_{\mathrm{tot}}$ is the total sum of squares, as defined by Equation~(3).

As an indicator of how good the model fits the real data, the $R^{2}$ score ranges from $0$ to $1$. The best possible score is $1$, and it can be negative if the model performs arbitrarily worse than a baseline. A constant model that always predicts the expected value of $y$, regardless of the input features, would obtain an $R^{2}$ score of $0$.

The evaluation of the model’s accuracy is completed by generating scores based on a separate data set from the one that was used for learning the parameters of prediction function for avoiding over–fitting. Instead of performing tests on the trained regression model with the held-out part of data called testing data set, the prediction scores in this study are obtained by using k-fold cross validation. The available data set is split into k smaller sets. The following procedure is followed for each of the k folds: (1) A model is trained using k-1 of the folds as training data; (2) The resulting model is validated on the remaining part of the data to give a performance measure such as score in this case; (3) The score reported by k-fold cross-validation is then the average of the values computed in the loop.

Although this approach need more computational resource than the conventional method, it enhances the accuracy of the model by avoiding the potential biases caused by an arbitrary test set while minimizing reduction of the size of training data set. In this case, k is taken as 5. It means the available data set was split into 5 smaller sets, each of which takes turns to be used as training data set while the rest of the data used as validation set. In order to save as much computation time as possible, while ensuring enough prediction accuracy, we use random permutation function to pick up 1000 data as a sample for one experiment (modeling training and validation) and repeat the process 10 times to conduct 10 experiments with different samples. In this way, the sample size reduces significantly compared to the original data set containing 56791 data points. The prediction score is computed by averaging the scores resulted from the 10 experiments with random picked samples to keep the test impartial.

\subsection{Feature selection}

In this research, there are 93 possible independent variables. Other than 5 time and date parameters (hour, day, weekday, month and year), 34 parameters regarding weather information and 11 parameters of historical TTI, the rest 43 variables are generated by transforming the date and time information from numerical into indicator variables. On one hand, there exist many redundant and/or insignificant information within the variable sets. On the other hand, it is tremendously time and space-consuming to deal with so many x-variables, especially when polynomial features are also involved. It is obvious that building regression model based on the whole set of variables is neither necessary nor feasible.

In order to identify and remove as much irrelevant independent variables as possible from the training data set, an algorithm called Recursive Feature Elimination (RFE) is executed in the data preparation of this study. As a feature selection algorithm, RFE follows a procedure that have three components. First, instead of exhaustively proposing subsets of features and attempting to find an optimal subset from them, RFE uses a backward stepwise selection algorithm which starts with all attributes in the set and gradually removes them one at a time. Second, an external estimator is used to assign coefficients to the proposed features and to evaluate the accuracy of the model. Third, a stopping criterion, which is that the addition or deletion of any feature will not produce a better subset (Karagiannopoulos et al., 2007), has to be met to stop iterations and establish the optimal solution. The procedure is recursively repeated on the pruned set until the desired number of features to select is eventually reached.

In this study, linear regression model is defined as external estimator. The desired number of features is iterated from 1 to 24. In other word, 24 optimal subsets at different sizes (1 to 24 elements) are tested conjunctively with different other factors (model, parameter, polynomial degree, etc.). The final model is selected out of those combinations by comparing their cross-validation scores.

\subsection{Model Comparison}

In order to find out the best model to predict TTI in DC area with high accuracy, 5 regression models provided by Scikit-learn \cite{Pedregosa2011ScikitLearn} are tested for performance comparison, including: linear regression, Lasso regression, Ridge regression, support vector machines–SVR and decision tree regressor. For the regression models whose accuracy might be sensitive to one or more parameters, we iterate each of the parameters within the pre-determined range based on the restrictions of algorithm and nature of the problem. For each regression model with different parameter values, it runs with optimized combination of independent variables selected by the feature selection algorithm.

Also, polynomial features are taken into consideration. Feature matrices consisting of all polynomial combinations of the features with degree less than or equal to the specified maximum degree are tested. For example, if the specified maximum degree of the input sample is $2$ and of the form $[a,\, b]$, the degree-2 polynomial features are $[1,\, a,\, b,\, a^{2},\, ab,\, b^{2}]$. The maximum degree takes its value in the range from $1$ to $5$.

In every single case, as a combination of different regression model, parameter value, independent variable set, and polynomial degree, the predictions are repeated for 10 times based on 1000 data points randomly selected from the historical TTI data. The final score in each case is calculated by averaging the resulted scores from the 10 calculations. 

Two cases are evaluated separately based on different time scales of historical data: (1) Case 1: historical TTI data used are up to 1 hour before the data point under prediction which is called short-term prediction case herein-below; (2) Case 2: historical TTI data up to 1 day before is called long-term prediction case. The best cases of short-term prediction case for each model are output and summarized in Table \ref{tab:tab2}. The results from long-term predictions are summarized in Table \ref{tab:tab3}.

\begin{table}[t]
\centering
\caption{Best prediction results and corresponding parameters and number of variables used in the prediction based on historical data up to 1 hour before.}
\label{tab:tab2}
\setlength{\tabcolsep}{6pt}
\renewcommand{\arraystretch}{1.15}
\begin{tabular}{l l c c}
\hline
\textbf{Model} & \textbf{Best parameters} & \textbf{\# Variables} & \textbf{Best score ($R^{2}$)} \\
\hline
Ridge Regression & $\alpha=1.0$ & 24 & 0.9025 \\
Linear Regression & -- & 21 & 0.8839 \\
SVR & $C=2.8,\ \epsilon=0.1$ & 10 & 0.8322 \\
Decision Tree Regressor & $\mathrm{max\_depth}=2.2$ & 21 & 0.6824 \\
Lasso Regression & $\alpha=0.1$ & 19 & 0.4265 \\
\hline
\end{tabular}
\end{table}

\begin{table}[t]
\centering
\caption{Best prediction results and corresponding parameters and number of variables used in the prediction based on historical data up to 1 day before.}
\label{tab:tab3}
\setlength{\tabcolsep}{6pt}
\renewcommand{\arraystretch}{1.15}
\begin{tabular}{l l c c}
\hline
\textbf{Model} & \textbf{Best parameters} & \textbf{\# Variables} & \textbf{Best score ($R^{2}$)} \\
\hline
Ridge Regression & $\alpha=1.9$ & 13 & 0.6086 \\
Linear Regression & -- & 15 & 0.6041 \\
SVR & $C=1.6,\ \epsilon=0.1$ & 15 & 0.5802 \\
Decision Tree Regressor & $\mathrm{max\_depth}=2.2$ & 11 & 0.5367 \\
Lasso Regression & $\alpha=0.19$ & 15 & 0.6062 \\
\hline
\end{tabular}
\end{table}

As shown in Table \ref{tab:tab2} and Table \ref{tab:tab3}, ridge regression is the best model out of the five models we have tested to predict TTI in both short- and long-term prediction cases. In short-term prediction case, a best score of 0.9025 is achieved when the parameter alpha has a value of 1.0 and the best 21 independent variables are selected to train the second-order polynomial regression model. 

In long-term prediction case, however, the best score achieved is 0.6086. It is significantly lower than the score in short-term case. This difference is attributed to the candidacy of the historical TTI values 1 to 3 hours before as independent variables which can therefore be extrapolated to be an important factor influencing the TTI prediction. The best value of parameter alpha is 1.9. Totally 15 variables together with their second-order terms and cross-product terms are involved in the best polynomial regression model.

\section{CONCLUSION}
In this research, instead of using TTI data for only tens of days, TTI data collected for 6 years are being used to provide more comprehensive information. Other than that, more weather indexes are also being used than the previous works to offer the research team a better insight into the relationship between weather and TTI. The increased amount of data involved will significantly improve the reliability and accuracy of the result reached by the research. During the research, it is surprising to find that instead of minimum visibility, the mean visibility is the weather index that has a more prominent effect on TTI. It is also surprising to find that the Ridge regression model outperform the linear regression model which is widely used in the previous works. This improved model, together with the findings in weather indexes, will provide industries a better chance to predict their travel time and therefore the monetary resource that is required to invest into new projects.

In conclusion, this paper presents travel time index (TTI) prediction using historical data. Five regression models provided by Scikit-learn package are tested for performance comparison, including: linear regression, Lasso regression, Ridge regression, support vector machines–SVR and decision tree regressor. Ridge regression is the best model out of the five models to predict TTI in both short and long-term prediction cases. In short-term prediction case, a best score of 0.9025 is achieved when the parameter alpha has a value of 1.0 and the best 21 independent variables are selected to train the second-order polynomial regression model. In long-term prediction case, the best score achieved is 0.6086. For future study, effort may be directed towards incorporating accidents information as an additional independent variable to improve the prediction models. Also, the relationship between each weather index should also be study to provide more knowledge of indirect relationship between weather indexes and TTI.

\bibliographystyle{unsrtnat}  
\bibliography{references}

\end{document}